\documentclass[prl,twocolumn,showpacs]{revtex4}

\usepackage{graphicx}
\usepackage[latin1]{inputenc}

\newcommand{\eq}[1]{(\ref{#1})}

\newcommand{\SKIP}[1]{}

\begin{document}

\title{Low Temperature Lanczos Method}

\author{Markus Aichhorn, Maria Daghofer, Hans Gerd Evertz, and Wolfgang von der Linden}

\affiliation{Institut f\"ur Theoretische Physik,
Technische Universit\"at Graz,
Petersgasse 16, A-8010 Graz, Austria}

\date{November 8, 2002}

\begin{abstract}
We present a modified finite temperature Lanczos method for the
evaluation of dynamical and static quantities of strongly
correlated electron systems that complements the finite temperature
method (FTLM) introduced by Jakli\v{c} and Prelov\v{s}ek
for low temperatures. Together they allow accurate
calculations at any temperature with moderate effort. As an example
we calculate the static spin correlation function and the regular
part of the optical conductivity $\sigma^{reg}(\omega)$ of the
one dimensional Hubbard model at half-filling and show in detail
the connection between the ground state and finite temperature method.
By using Cluster Perturbation Theory (CPT), 
the finite temperature spectral function is extended to the infinite system,
clearly exhibiting the effects of spin-charge separation.
\end{abstract}


\maketitle

The Finite Temperature Lanczos Method (FTLM), 
introduced by Jakli\v{c} and Prelov\v{s}ek \cite{JaPr94},
has in recent years allowed the
precise calculation
of thermodynamic quantities of strongly correlated systems.
It has been applied to
the $t$-$J$ model for the cuprates \cite{JaPr95,JaPr95_2,JaPr96,PrRa00,VePr00,ShTo01,VePr02} and 
vanadates \cite{CuHo99,AiHo02}, orbital $t$-$J$ model \cite{HoJa99}, Kondo
lattice model \cite{HaBo00}, Heisenberg
model \cite{ScYu02} and static properties of the Hubbard
model \cite{Hell01,BoPr02}. In principle this method can be
applied at all temperatures, but at low temperatures the required number
of random samples is very large.
FTLM is restricted to small systems and particularly at low temperatures,
finite size effects become important. 
They can be overcome, at least for properties derived from the single
particle Green's function, by using Cluster Perturbation Theory (CPT)
\cite{SePe00,SePe02}, 
which provides infinite system results with remarkable accuracy \cite{Inprep}.
However, CPT requires finite cluster Greens functions $G_{ab}(\omega)$ for
all sites $a,b$, increasing the number of required Lanczos runs and matrix
elements by a 
factor equal to the number of lattice sites. It therefore requires a
sufficiently fast low temperature method.

In this paper we present a modified finite temperature Lanczos method which
allows to calculate properties for large Hilbert spaces at low temperatures
that are not accessible by 
the existing method. We show that a combination of our low temperature
Lanczos method (LTLM) with the FTLM allows an accurate calculation of thermodynamic
properties at any temperature with moderate effort.

Let us first present the existing FTLM. For the case of a static expectation
value of an operator ${\mathcal{O}}$
\begin{equation}\label{eq1}
  \langle{\mathcal{O}}\rangle=\frac{1}{Z}\sum_n^N\langle n|{\mathcal{O}}e^{-\beta
  H}|n\rangle,\quad Z=\sum_n^N\langle n|e^{-\beta
  H}|n\rangle,
\end{equation}
with $\beta=1/T$ ($k_B=\hbar=1$) and a sum over a complete orthonormal basis set $|n\rangle$,
the FTLM approximation is 
\begin{eqnarray}
  \langle{\mathcal O}\rangle &=& \frac{1}{Z}\sum_s\frac{N_s}{R}\sum_r^R\sum_m^Me^{-\beta
  \varepsilon_m^{(r)}}
  \langle r|\Psi_m^{(r)}\rangle\langle\Psi_m^{(r)}|{\mathcal
  O}|r\rangle,\nonumber\\
  Z &=& \sum_s\frac{N_s}{R}\sum_r^R\sum_m^M|\langle r|\Psi_m^{(r)}\rangle|^2e^{-\beta
  \varepsilon_m^{(r)}},\label{eq2}
\end{eqnarray}
with a sum over symmetry sectors $s$ of dimension $N_s$,
and $R$ random vectors $|r\rangle$ in each sector \cite{comment}. 
$M$ is the number of Lanczos steps. 
For each random vector $|r\rangle$, a Lanczos procedure is performed, 
yielding $M$ eigenenergies $\varepsilon_m^{(r)}$ and corresponding eigenvectors $|\Psi_m^{(r)}\rangle$. 
The trace in \eq{eq1} requires $N_s$ states, while 
very accurate results can be obtained via \eq{eq2} even for a drastically
reduced number of Lanczos steps $M\ll N_s$ and a partial random sampling of the
Hilbert subspaces $R\ll N_s$.

For dynamical correlation functions $C(t)=\langle A(t)B\rangle$,
FTLM calculates
\begin{eqnarray}
  C(t) &=&  \frac{1}{Z}\sum_s\frac{N_s}{R}\sum_r^R\sum_{i,j}^M e
  ^{-\beta\varepsilon_i^{(r)}} 
  e^{-i(\tilde\varepsilon_j^{(r)}-\varepsilon_i^{(r)})t} \times\nonumber\\
  &\times&\langle r|\Psi_i^{(r)}\rangle\langle\Psi_i^{(r)}|A|\tilde\Psi_j^{(r)}\rangle\langle\tilde\Psi_j^{(r)}|B|r\rangle,
  \label{eq4}
\end{eqnarray}
Here, an initial vector
\begin{equation}\label{eq5}
  |\tilde\Phi_0^{(r)}\rangle=B|r\rangle/\sqrt{\langle r|B^\dagger B|r\rangle}.
\end{equation}
is used to generate additional eigenenergies  $\tilde\varepsilon_j^{(r)}$ 
and eigenvectors $|\tilde\Psi_j^{(r)}\rangle$ from 
that part of the Hilbert space onto which the operator $B$ projects.
Hence the term
$\langle\tilde\Psi_j^{(r)}|B|r\rangle$ in \eq{eq4} becomes sufficiently large.

Let us now consider the behavior of (\ref{eq2}) and (\ref{eq4}) in the limit $T\to 0$. 
In this case only the ground state $|\Psi_0\rangle$
contributes and we get
\begin{equation}
  \langle{\mathcal O}\rangle=\sum_r^R\langle\Psi_0|{\mathcal
  O}|r\rangle\langle r|\Psi_0\rangle /  \sum_r^R\langle\Psi_0|r\rangle\langle
  r|\Psi_0\rangle 
\end{equation}
and similarly for \eq{eq4}. 
\SKIP{The exact ground state result $\langle\Psi_0|{\mathcal O}|\Psi_0\rangle$ can
therefore only be 
approached if $R\sim N_{s=0}$, in other words sampling over the whole Hilbert
subspace containing the ground state, because then the sum over $|r\rangle$
reduces to the identity 
operator. If one uses only few random sampling vectors, the statistical errors
become very large, a severe limitation to the applicability of the
method. This limitation extends to small finite temperatures, where only few states
contribute to the partition function.}
Thus the ground state result will suffer from severe statistical fluctuations,
although the exact (Lanczos) eigenvector $|\Psi_0\rangle$ is reached with
every $|r\rangle$ and one random vector should be sufficient. Yet,
FTLM gets worse with decreasing temperature $T$.

The modifications we present in this paper are designed to overcome this limitation.
Let us introduce the method for a static expectation value \eq{eq1}. 
We use a symmetric form
\begin{equation}\label{eq7}
  \langle{\mathcal O}\rangle=\frac{1}{Z}\sum_n^N\langle
  n|e^{-\frac{1}{2}\beta H}{\mathcal O}e^{-\frac{1}{2}\beta H}|n\rangle.
\end{equation}
As before, we approximate the trace by random
sampling, but now we insert the approximate eigenbasis obtained by the
Lanczos procedure twice, initially obtaining
\begin{eqnarray}
  \langle{\mathcal
  O}\rangle&=&\frac{1}{Z}\sum_s\frac{N_s}{R}\sum_r^R\sum_{i,l}^Me^{-\frac{1}{2}\beta(\varepsilon_i^{(r)}+
  \varepsilon_l^{(r)})}\times\nonumber\\
  &\times&\langle r|\Psi_l^{(r)}\rangle 
  \langle\Psi_l^{(r)}|{\mathcal
  O}|\Psi_i^{(r)}\rangle\langle\Psi_i^{(r)}|r\rangle \label{eq8}
\end{eqnarray}
The partition function $Z$ is calculated in the same way as in standard FTLM. 
The behavior in the limit $T\to 0$ is now different. 
If only the ground state $|\Psi_0\rangle$ contributes, \eq{eq8} becomes
\begin{eqnarray}
  \langle{\mathcal O}\rangle&=&\sum_r^R\langle
  \Psi_0|r\rangle\langle r|\Psi_0\rangle\langle\Psi_0|{\mathcal
  O}|\Psi_0\rangle / \sum_r^R \langle\Psi_0|r\rangle\langle r|\Psi_0\rangle\nonumber\\
  &=&\langle\Psi_0|{\mathcal O}|\Psi_0\rangle.\label{eq9}
\end{eqnarray}
\SKIP{Here it is not necessary to have a large number of random samples, 
because the sum over $|r\rangle$ need not converge to the unity operator. 
By applying a jackknife analysis 
to the statistical data,
the terms $\langle r|\Psi_0\rangle$ which carry the statistical fluctuations
cancel and one recovers the
ground state result already for a
very small number of random samples, even for one random vector
$|r\rangle$ in the symmetry sector of the ground state.}
In agreement with ground state Lanczos, one random vector suffices for the
ground state expectation value.

If we compute the numerator in \eq{eq8} and
$Z$ separately, both suffer from pronounced statistical fluctuations, which however
cancel exactly at $T=0$ as shown in \eq{eq9}. For finite $T$ the fluctuations
in numerator and denominator do not cancel exactly but they are still
strongly correlated. Separate error analysis for both terms would overestimate the
statistical noise. These correlations are taken into account by employing a
Jackknife technique \cite{jackknife}.

For dynamical correlation functions, a straight-forward variant of \eq{eq4} 
suitable for low temperatures is
\begin{widetext}
\begin{equation}\label{eq10}
  C(t)=\frac{1}{Z}\sum_s\frac{N_s}{R}\sum_r^R\sum_{ijl}^M e^{-\frac{1}{2}\beta(\varepsilon_i^{(r)} +\varepsilon_l^{(r)})}
  e^{-i(\tilde\varepsilon_j^{(r)}-\frac{1}{2}(\varepsilon_i^{(r)}+\varepsilon_l^{(r)}))t}
  \langle
  r|\Psi_l^{r}\rangle\langle\Psi_l^{(r)}|A|\tilde\Psi_j^{(r)}\rangle
  \langle\tilde\Psi_j^{(r)}|B|\Psi_i^{(r)}\rangle
  \langle\Psi_i^{(r)}|r\rangle \,
\end{equation}
\end{widetext}
In order to span the relevant subspace of the Hilbert space,
we now choose initial vectors
$|\tilde\Phi_0^{(r,i)}\rangle\propto B|\Psi_i^{(r)}\rangle$
for the second Lanczos run.
With $M$ such second Lanczos runs, the numerical effort would be much higher than for FTLM.
For low temperatures, it can be reduced,
since only the low lying states contribute to the expectation values.
We consider only states below a cutoff energy $E_c$, defined by
\begin{equation}\label{eq11}
   e^{-\beta_{c}(E_{c}-E_0)}<\varepsilon_{c},
\end{equation}
where $\varepsilon_{c}$ defines the accuracy of the approximation, $\beta_c$
is the minimal inverse temperature considered and the calculation will be
accurate for all $\beta>\beta_c$. 
We thus proceed as follows:
For each random start vector $|r\rangle$, we perform
an initial Lanczos run with $M$ iterations.
For each of the $M_c$ states $|\psi_i^{(r)}\rangle$ with energies below $E_c$,
we then calculate an initial vector  
$|\tilde\Phi_0^{(r,i)}\rangle\propto B|\Psi_i^{(r)}\rangle$
and perform a second run with $M$ Lanczos iterations,
obtaining an approximate eigenbasis $|\tilde\Psi_j^{(r,i)}\rangle$.
Using these basis sets, the final form of LTLM 
is the same as \eq{eq8} and \eq{eq10},
with $\sum_{i,l}^M$ and $\tilde{\Psi}_j^{(r)}$
replaced by $\sum_{i,l}^{M_c}$ and $\tilde{\Psi}_j^{(r,i)}$, respectively.

Memory requirements of our method are the same as for standard
FTLM, but the CPU time requirements differ significantly.
CPU time is mainly determined by the number of matrix elements that have to
be calculated. In the case of static expectation values these are $M$ for
FTLM and $M_c^2$ for LTLM for each random vector. Therefore both methods reach
equivalent CPU time requirements per random vector when $M_c\approx\sqrt{M}$.

For dynamical correlation functions,
the number of matrix elements to be calculated in the second Lanczos run
is $M^2$ for FTLM and $M_c^2$ for LTLM. 
For LTLM we have to perform $M_c$ second Lanczos runs, but only one for FTLM.
Thus we have similar CPU 
time requirements per random vector for both methods when
$M_c\approx M^{\frac{2}{3}}$.
In the limit $T\to 0$ we have $M_c=1$, and for $R=1$, LTLM
is comparable to the ground state Lanczos technique.

For both methods, CPU-time is proportional to $R$, the number of random vectors.
But, by design, far less random vectors are
needed for the LTLM than for the FTLM at low temperatures
for a given accuracy.

All considerations so far have been done without regarding
reorthogonalization of Lanczos vectors. This procedure becomes important for $M\gtrsim
150$, where numerical round off errors become significant, and significantly
increases CPU requirements. 

Let us now demonstrate the method for the calculation of static and dynamical
properties of the one dimensional Hubbard model,
with hamiltonian
\begin{equation}\label{eq16}
  H=-t\sum_{i,\sigma}(c_{i\sigma}^\dagger
  c_{i+1,\sigma}+\mbox{h.c.})+U\sum_in_{i\uparrow}n_{i\downarrow}.
\end{equation}
We specify energies in units of $t$.
As an example 
we calculate the static spin
correlation function $C_1=\langle S_i^zS_{i+1}^z\rangle$ on a 12 site chain
with  periodic boundary conditions at half filling ($n=1$). 
The number of basis states is $N=$~2~704~156. 
Symmetry sectors are specified by 
momentum $\mathbf{k}$ and total spin $S_z$.
The largest sector $S_z=\pm 1$ has 52~272 basis states. The sector $S_z=0$ is
further reduced due to spin up/down symmetry. 

\begin{figure}[htb]
  \centering
  \includegraphics[width=3.2in]{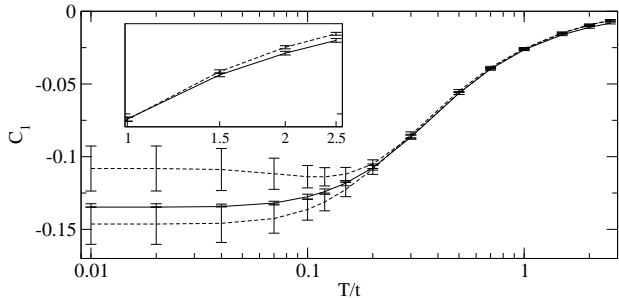} 
  \caption{\label{fig1}%
    Spin correlation function $C_1=\langle S_i^zS_{i+1}^z\rangle$ for the
    1D Hubbard model on a 12 site chain with periodic boundary conditions at U=8t and n=1. 
    Solid: LTLM with  $\beta_c=1$, $\varepsilon_c=0.01$, $M=100$, and  $R=25$.
    Dashed: Two independent runs of FTLM with $M=100$ and $R=25$. 
    Inset: Deviation of $C_1$ in the high temperature region beyond $\beta_c$.
    Here $R=50$ in both cases.}
\end{figure}
\begin{figure}[htb]
  \centering
  \includegraphics[width=2.9in]{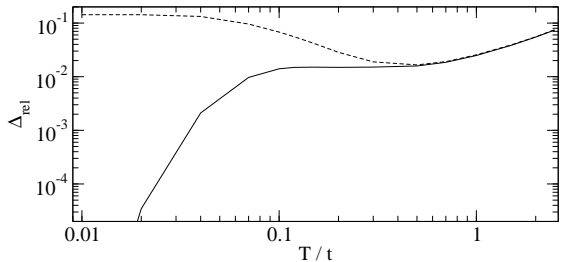}
  \caption{\label{fig1_1}%
     Relative statistical errors $\frac{\Delta C_1}{C_1}$
     of LTLM (solid) and FTLM (dashed) with $R=25$ in both cases.
     Other parameters as in fig.\ \ref{fig1}. 
     The error of LTLM at $T=0.01t$ is $\Delta_{rel}=10^{-8}$.}
\end{figure}  

In figs. \ref{fig1} and \ref{fig1_1} the convergence and statistical errors  
of LTLM and FTLM are compared
at equal computational effort, 
with $R=25$ random samples per symmetry sector each,
corresponding to sampling of 
$\frac{R}{N_s}\approx 0.05\%$ of the largest Hilbert subspace.
At low temperatures, our method provides results which are orders of magnitude 
more precise than from standard FTLM,
and which connect smoothly to the ground state properties.
We checked that for larger $R$, there is no systematic drift for either method,
and the FTLM results converge towards those of LTLM.
At intermediate temperatures, the statistical errors of LTLM increase and become 
similar to those of FTLM.
Finally, considerably beyond the chosen cutoff-temperature $1/\beta_c$,
LTLM is no longer applicable,
and begins to show a systematic deviation.

Both FTLM and LTLM provide results for a range of 
temperatures from a single calculation.
For FTLM this range is limited towards low temperatures by statistical errors.
For LTLM, it is limited by the chosen cutoff-temperature $1/\beta_c$.
Therefore a combination of both methods provides precise results
for all temperatures with moderate effort.

\begin{figure}[t]
  \centering
  \includegraphics[width=3.2in,height=3in]{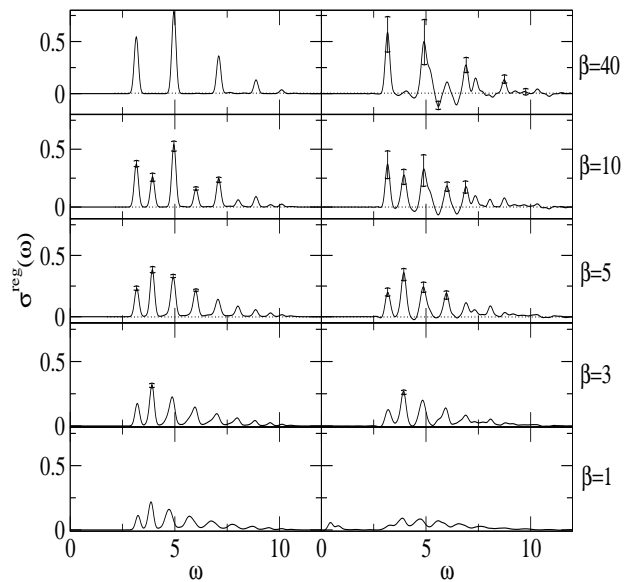}
  \caption{\label{fig2}%
    Regular part of the optical conductivity of the 1d Hubbard
    model on a 12 site chain with periodic boundary conditions at U=6t and n=1. 
    Left panel: LTLM calculations with
    $\beta_{c}=3$, $\varepsilon_{c}=0.01$, and $R=40$. Right panel: FTLM
    calculations with
    $R=50$. Number of Lanczos steps $M=100$ and additional broadening of
    $\sigma=0.1$. 
    Dots mark the zero line. 
    Only selected errorbars are shown.
    For curves without errorbars, the errors are smaller than the line width.} 
\end{figure}

As an example of
dynamical correlation functions
we calculate the regular part
of the optical conductivity, given by the current-current correlation function
\begin{equation}
  \sigma^{reg}=\frac{1-e^{-\beta\omega}}{\omega}\mbox{Re}\int_0^\infty\!\!dt\,e^{i\omega
  t} \langle j(t)j\rangle,
\end{equation}
with the current operator
  $ j=it\sum_{i,\sigma}(c_{i\sigma}^\dagger  c_{i+1,\sigma}-\mbox{h.c.})$.
In fig. \ref{fig2}, we show results with 
approximately the same CPU time for both methods.
Slightly above the ground state, at $\beta=40$, 
LTLM approaches the exact ground state
result \cite{FyMa91,JeGe00}.
For intermediate temperatures $\beta=10,5,3,$ slight
statistical fluctuations occur. 
By comparison to FTLM we see that $\beta=1<\beta_c$ is indeed beyond the
validity of this calculation. We also checked the accuracy
of the results by using $M=200$ Lanczos steps instead of $M=100$ yielding the
same LTLM spectra within statistical errors.

In contrast, FTLM suffers from
strong statistical fluctuations at small temperatures.
Errorbars are very large and regions occur where
$\sigma^{reg}(\omega)$ 
becomes negative, a clear indicator that we did not use enough random vectors for
FTLM. As expected from our consideration of static expectation values,
errorbars of FTLM get smaller for higher temperatures. 
As for LTLM we did calculations with
$M=200$, yielding the same curves within errorbars but leading to a better
convergence at the high frequency side of the 
spectrum.

\begin{figure}[ht]
  \centering
  \includegraphics[width=3.2in,height=3.6in]{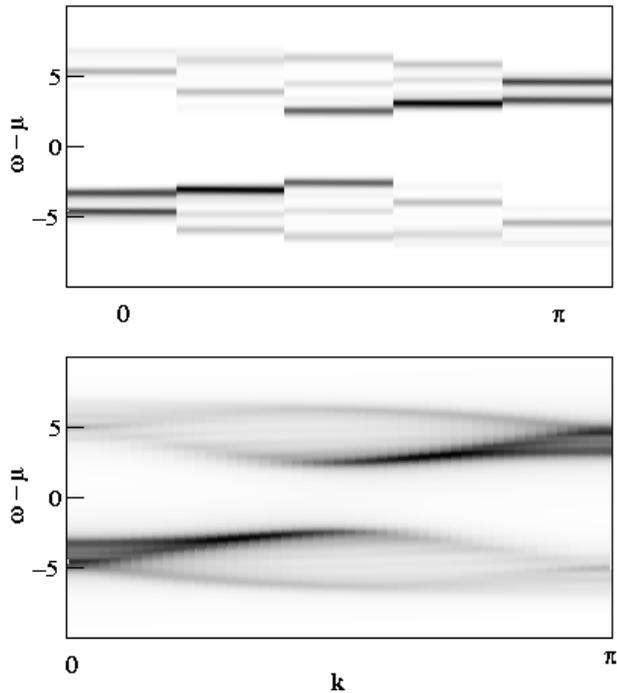}
  \caption{\label{fig4}%
    Spectral function $A(\mathbf{k},\omega)$ obtained by LTLM for the 1D Hubbard model at
    $U=8t$, $n=1$, and $\beta=10$. Parameters: $R=30$, $\beta_c=3$,
    $\varepsilon_c=0.01$. Upper panel: 8 site chain with periodic boundary conditions. 
    Lower panel: CPT result based  on 8 site clusters. 
    }
\end{figure}  

As mentioned in the beginning, at low temperature finite size effects become
important. At least for properties derivable from the single particle Green's
function, these effects can be overcome by using CPT \cite{SePe00,SePe02}. In
fig. \ref{fig4} we show the finite temperature spectral function
$A(\mathbf{k},\omega)$ obtained on a 
finite size lattice with periodic boundary conditions,
and the infinite lattice result obtained by CPT, 
which makes use of all Greens functions on the finite lattice as calculated by LTLM.
In the latter a smooth structure consisting of several branches
can clearly be seen with spin-charge separation at $\mathbf{k}=0$ visible in
the lower part of the spectrum \cite{SePe00,ZaAr98}. On the finite size cluster,
however, this structure is not evident
as it exhibits more discrete excitations.
Further work on finite temperature CPT is in progress \cite{Inprep}.

In conclusion, the method presented in this paper gives an accurate
connection of the exact ground state Lanczos method and the 
established FTLM.
Using LTLM at low and FTLM at higher temperatures makes it possible to
calculate static and dynamical properties of 
strongly correlated systems from $T=0$ up to $T=\infty$ with very good
accuracy and rather small numerical effort. 

This work has been supported by the Austrian Science Fund (FWF)
projects P15834 and P15520.
M. Aichhorn is supported by DOC [Doctoral Scholarship Program of the Austrian
Academy of Sciences].


\begin{thebibliography}{99}
\bibitem{comment} In standard FTLM, $R \sim N_s$ in each sector.
                    Since the computational effort per random vector is largest in
                    the biggest symmetry sector, \eq{eq2} provides a better
                    sampling of smaller sectors at negligible additional cost.
\end{thebibliography}
\end{document}